\documentclass[fleqn,twoside]{article}
\usepackage{espcrc2}
\usepackage{graphicx}
\usepackage[figuresright]{rotating}

\newcommand{\AmS}{{\protect\the\textfont2
  A\kern-.1667em\lower.5ex\hbox{M}\kern-.125emS}}

\title{The Search for Anisotropy in the Arrival Directions of Ultra-High 
Energy Cosmic Rays Observed by the High Resolution Fly's Eye Detector in 
Monocular Mode}

\author{Benjamin~T.~Stokes
	\address[utah]{University of Utah, Department of Physics
	and High Energy Physics Institute,\\ 
        115 S 1400 E RM 201, Salt Lake City, UT 84112-0830, USA\\
	stokes@cosmic.utah.edu}
	for the High Resolution Fly's Eye (HiRes) Collaboration
       	\thanks{See http://hires.phys.columbia.edu for a complete list of
	authors}}
\begin{document}

\begin{abstract}
The High Resolution Fly's Eye HiRes-I detector has now been in 
operation in monocular mode for over six years.  During that time span, 
HiRes-I has accumulated a larger exposure to Ultra-High Energy Cosmic Rays 
(UHECRs) above $10^{19}$~eV than any other experiment built to date.  
This presents an unprecedented opportunity to 
search for anisotropy in the arrival directions of UHECRs.  
We present results of a search for dipole distributions oriented towards 
major astrophysical landmarks and a search for small-scale clustering. 
We conclude that the HiRes-I data set is, in fact, consistent with an isotropic
source model.

\vspace{1pc}
\end{abstract}

\maketitle

\section{Introduction}

The observation of Ultra-High Energy Cosmic Rays (UHECRs) has now spanned 
over forty years.  Over that period, many source models have been 
proposed to explain the origin of these remarkable events. In the past five 
years, theoretical models have been suggested that would potentially produce 
dipole distributions oriented towards M87 \cite{Biermann:fd} or
Centaurus A \cite{Farrar:2000nw,Anchordoqui:2001nt}.  In addition, the 
Akeno Giant Air Shower Array (AGASA) has reported findings
suggesting a 4\% dipole-like enhancement oriented towards the Galactic Center 
present in its events with energies around 
$10^{18}$~eV \cite{Hayashida:1999ab}.  This result seemed to be
corroborated by findings published by the Fly's Eye experiment in 1999 that  
suggested the possibility of an 
enhancement in the galactic plane also at energies around $10^{18}$~eV
 \cite{Bird:1998nu}, and also
by a re-analysis of data from the SUGAR array that was published in 2001 
\cite{Bellido:2000tr} that showed an enhancement in the general vicinity of
the Galactic Center.  

However, both AGASA and Fly's Eye are subject to a limiting factor; 
they are both located too far north in latitude to directly 
observe the Galactic Center 
itself.  The re-analysis of SUGAR data actually demonstrated an excess 
that was offset from the Galactic Center by $7.5^\circ$ and was more consistent
with a point source than a global dipole effect \cite{Bellido:2000tr}.
While the current High Resolution Fly's Eye (HiRes) experiment 
is subject to a similar 
limitation in sky coverage as the AGASA and Fly's Eye experiments, 
we will show that, by properly estimating the HiRes aperture and
angular resolution, we can effectively exclude these dipole source  
models to a certain degree of sensitivity.  However, we are not able to 
completely exclude the findings of AGASA or the theoretical predictions
mentioned above.

Additionally, over the past decade, 
the search for sources of Ultra-High Energy 
Cosmic Rays (UHECRs) has also begun to focus upon small scale anisotropy 
in event arrival directions.  
This refers to statistically significant excesses occurring at the
scale of $\leq2.5^\circ$.  The interest in this sort of anisotropy has
largely been fueled by the observations of the Akeno Giant Air Shower Array 
(AGASA).  In 1999 \cite{Takeda:1999sg} and again in 
2001 \cite{Takeda:2001}, the AGASA collaboration reported observing
what eventually became seven clusters (six ``doublets'' and one ``triplet'') 
with estimated energies above $\sim3.8\times10^{19}$~eV.  
Several attempts that have been made to 
ascertain the significance of these clusters returned chance probabilities
of $4\times10^{-6}$ \cite{Tinyakov:2001ic} to 0.08 
\cite{Finley:2003on}.

By contrast, the monocular (and stereo) analyses that have been 
presented by the High Resolution Fly's
Eye (HiRes) demonstrate that the level of autocorrelation observed 
in our sample is completely consistent with that expected from background
coincidences \cite{bellido,bellido2,apj}.  
Any analysis of HiRes monocular data needs to take into account that the 
angular resolution in monocular mode is highly asymmetric.  

It is very difficult to compare the results of the HiRes
monocular and AGASA analyses.  They are 
very different in the way that they measure autocorrelation.  
Differences in the published energy spectra of the two experiments suggest
an energy scale difference of 30\% \cite{prl,Takeda:1998ps}.
Additionally, the two experiments observe UHECRs in very 
different ways.  The HiRes experiment has an energy-dependent aperture and
an exposure with a seasonal variability \cite{prl}.  These differences
make it very difficult get an intuitive grasp of what HiRes should see if
the AGASA claim of autocorrelation is justified.   In order to develop this
sort of intuition, we apply the same analysis to both AGASA and HiRes data.  

Our methods for detecting the presence of a dipole source model and 
small-scale clustering are
based upon comparisons between the real data and a large quantity of 
events generated by our Monte Carlo
simulation program.  The simulated data possess
the same aperture and exposure as the actual HiRes-I monocular data set.
We show how the asymmetric angular resolution of a monocular air
fluorescence detector can be accommodated in this method.

\section{The HiRes-I Monocular Data}

The data set that we consider consists of events
that were included in the HiRes-I monocular spectrum measurement 
\cite{prl,fadc}.
This set contains the events observed between May~1997 and 
February~2003.  There were 1526 events with reconstructed energies above
$10^{18.5}$~eV and 52 events with reconstructed energies above
$10^{19.5}$~eV observed during this time period. 
The data set represents a cumulative 
exposure of $\sim3000$~km$^2\cdot$sr$\cdot$yr at $5\times10^{19}$~eV.
This data was subject to a number of quality cuts that are detailed in the
above-mentioned papers \cite{prl,fadc}.  We previously verified that this data
set was consistent with Monte Carlo predictions in many ways including
 impact parameter 
($R_p$) distributions \cite{prl} and zenith angle distributions \cite{dipole}. 
For this study, we presumed an average atmospheric clarity
\cite{Wiencke:atmos}.

\begin{figure}[t,b]
\begin{tabular}{c}
\includegraphics[width=5.5cm]{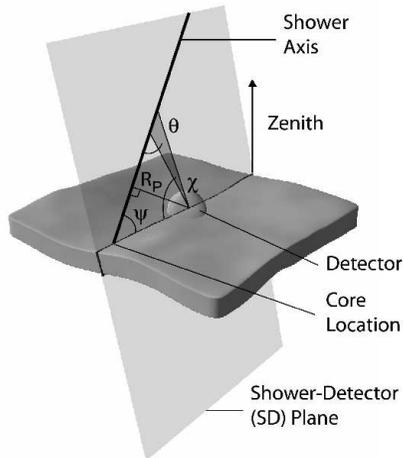}\\
\end{tabular}
\caption{The geometry of reconstruction for a monocular air fluorescence 
detector}
\label{figure:picture}
\end{figure}

In order to perform anisotropy analysis on this subset of data,
we must first parameterize the HiRes-I monocular angular resolution.
For a monocular air fluorescence detector, angular resolution consists
of two components, the plane of reconstruction, that is the plane in which
the shower is observed, and the angle $\psi$ within
the plane of reconstruction (see figure~\ref{figure:picture}). 
We can determine the plane of reconstruction 
very accurately.  However, the value 
of $\psi$ is more difficult to determine accurately because it is
dependent on the precise results of the profile-constrained fit 
\cite{prl,fadc}.

\begin{figure*}[t,b]
\begin{tabular}{c}
\includegraphics[width=12.0cm]{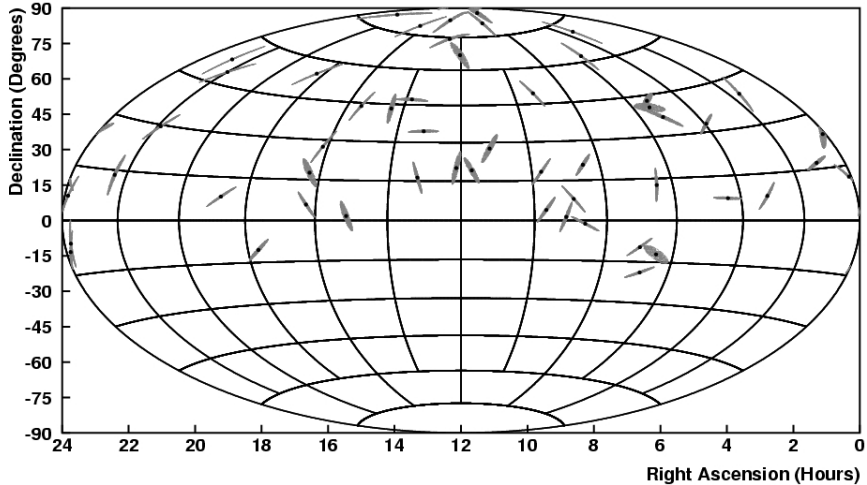}\\
\end{tabular}
\caption{The arrival directions of the HiRes-I monocular data 
with reconstructed 
energies above $10^{19.5}$~eV events and their $1\sigma$ angular resolution}
\label{fig:hires_map}
\end{figure*}

The HiRes-I angular resolution is therefore 
described by an elliptical, two-dimensional 
Gaussian distribution with the two Gaussian parameters, $\sigma_\psi$ and 
$\sigma_{\rm plane}$, being defined by the two angular resolutions.
For events with reconstructed energies above $10^{19.5}$~eV, 
$\sigma_\psi=[4.9,6.1]^\circ$ and $\sigma_{\rm plane}[0.4,1.5]^\circ$.
In figure~\ref{fig:hires_map}, 
the arrival directions of the
HiRes-I events  are plotted in
equatorial coordinates along with their $1\sigma$ error ellipses.

\begin{figure}[t,b]
\begin{tabular}{c}
\includegraphics[width=7.0cm]{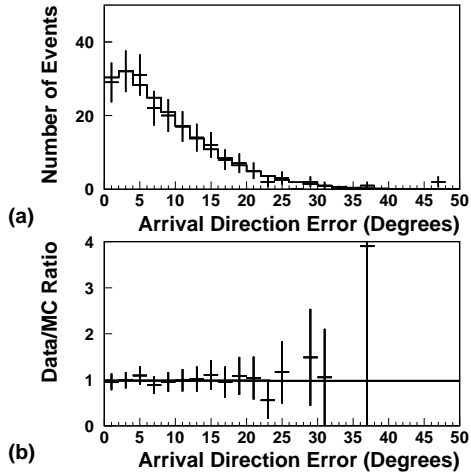}\\
\end{tabular}
\caption{Arrival direction error comparison between real data (mono vs. stereo)
and simulated data for events with estimated energies above $10^{18.5}$~eV.
The solid line histogram corresponds to the 
arrival direction error distribution of the monocular reconstructed 
Monte Carlo simulated data.  
The crosses correspond to the arrival directions error 
distribution observed for actual data by comparing the arrival directions 
estimated by the monocular and stereo reconstructions.  The solid line in the 
ratio component corresponds to the fit $y=ax+b$ where $a=0.000\pm0.011$ and
$b=0.98\pm0.11$.}
\label{fig:comp}
\end{figure}

In order to understand the systematic uncertainty in the angular resolution 
estimates, we 
consider a comparison of estimated arrival directions that successfully
reconstructed in both HiRes-I monocular mode and HiRes stereo mode.  
We consider 
all mono/stereo candidate events with estimated energies above $10^{18.5}$~eV.
In stereo mode, the shower detector planes of the two detectors are 
intersected, thus the geometry is much more precisely known and the total
angular resolution is of order $0.6^\circ$, a number that is
largely correlated to $\sigma_{\rm plane}$ and thus is negligible
when added in quadrature to the larger term, $\sigma_\psi$.  This allows us
to perform a comparison of the angular resolution estimated through simulations
to the observed angular resolution values of actual data.
In figure~\ref{fig:comp},
we show the distribution of angular errors for real and simulated data.  
The uncertainty in the slope of the ratio (figure~\ref{fig:comp}b) leads to 
an 7.5\% uncertainty in the angular resolution.

\section{Dipole Measurement Results}

The method for measuring the anisotropy amplitude, $\alpha$, 
for potential dipole sources 
is discussed at length in Abbasi {\it et al.} \cite{dipole}. 

\subsection{Analysis}

In summary, we measure the value of the anisotropy amplitude by the 
following method:

\begin{enumerate}
\item We calculated the value of $<\!\!\cos\theta\!\!>$ for the dipole function of the real data sample.  
\item We created a total of 20,000 simulated data samples, 1000 each for 
0.1 increments of $\alpha$ from -1.0 to 1.0, each with the same number of 
events as the actual data.   
\item We constructed curves corresponding to the mean and standard
deviation of $<\!\!\cos\theta\!\!>$ of the dipole function for each value of $\alpha$. 
\item We determined the preferred value of $\alpha$ and the 90\% 
confidence interval of $\alpha$ for each dipole source model by referring to 
the intersections of the 90\% confidence interval curves with the actual value
of $<\!\!\cos\theta\!\!>$ for the dipole function of the real data.
\end{enumerate}

The results are shown in table~\ref{table:comp}.

\begin{table}[t,b]
\begin{tabular}{|c|c|} \hline
{\bf Source} & $\alpha$ \\ \hline\hline
Galactic & $0.005\pm0.055$ \\ \hline
Centaurus A & $-0.005\pm0.065$ \\ \hline
M87 & $-0.010\pm0.045$ \\ \hline
\end{tabular}
\caption{Estimation of $\alpha$ via the value of $<\!\!\cos\theta\!\!>$ 
for the dipole function.}
\label{table:comp}
\end{table}

\subsection{Conclusion}

We are able to place upper limits on the value of $|\alpha|$ 
for each of our three proposed dipole source models.  However, these limits 
are not small enough to exclude the theoretical predictions 
\cite{Biermann:fd,Farrar:2000nw,Anchordoqui:2001nt}.  Also,
they do not exclude the findings of the AGASA collaboration in terms of the
intensity of the dipole effect that they observed or in terms of the energy
considered because the events in the 
dipole effect observed by the AGASA detector
possessed energies below $10^{18.5}$~eV \cite{Hayashida:1999ab}.  
Since it appears that angular resolution has little impact on the measurement
of $\alpha$ and we do not appear to be systematically limited,  
we conclude that the driving factor in making a better 
determination of $\alpha$ will simply be larger event samples.  HiRes-I
mono will continue to have the largest cumulative aperture of any single 
detector for the next three to five years, thus
it will continue to serve as an ever
more powerful tool for constraining dipole source models.

\section{Small-Scale Clustering Results}

The method for measuring the small scale clustering is discussed at length
in Abbasi {\it et al.} \cite{autocorr}.  

\subsection{Analysis}

We measure the degree of small-scale clustering by means
of an autocorrelation function.  It is calculated as follows:
\begin{enumerate}
\item For each event, an arrival direction is sampled on a probabilistic basis
from the error space defined by the  angular resolution of the event.
\item The opening angle is measured between the arrival directions of a pair
of events.
\item The cosine of the opening angle is then histogrammed.
\item The preceding steps are repeated until all possible pairs of the 
events are considered.
\item The preceding steps are repeated until the error space, in the arrival
direction of each event, is thoroughly sampled.
\item The histogram is normalized and the resulting curve is the 
autocorrelation function.
\end{enumerate}

A well-behaved measure of the 
autocorrelation of a specific set of data is the value of
$<\!\!\cos\theta\!\!>$ for $\theta\leq10^\circ$.    
This value is also a measure of the 
sharpness of the autocorrelation peak at $\cos\theta=1$.  However, this 
method of quantification  does not
depend on bin width and it does produce Gaussian distributions 
when it is applied to large numbers of
sets with similar degrees of autocorrelation.
An additional advantage to this method is that by considering the continuous
autocorrelation function over a specified interval, both the peak at 
the smallest values of 
$\theta$ and the corresponding statistical deficit in the autocorrelation
function at slightly higher values
of $\theta$ are taken into account.  Thus we simultaneously measure both the
positive and negative aspects of the autocorrelation signal.

\begin{figure}[t,b]
\begin{tabular}{c}
\includegraphics[width=5.9cm]{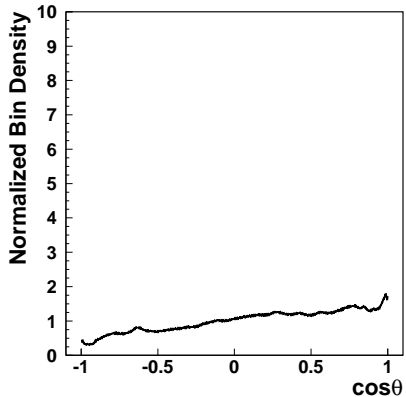}\\
\end{tabular}
\caption{The autocorrelation for the HiRes-I events above $10^{19.5}$~eV.}
\label{fig:hires_au}
\end{figure}

Using the description of the HiRes-I monocular angular 
resolution above, we then calculate the autocorrelation function via the 
method described above.  In figure~\ref{fig:hires_au}, 
we show the result of this calculation.  For this sample, we obtain 
$<\!\!\cos\theta\!\!>_{[0^\circ,10^\circ]}=0.99234$.  

\begin{figure}[t,b]
\begin{tabular}{c}
\includegraphics[width=5.9cm]{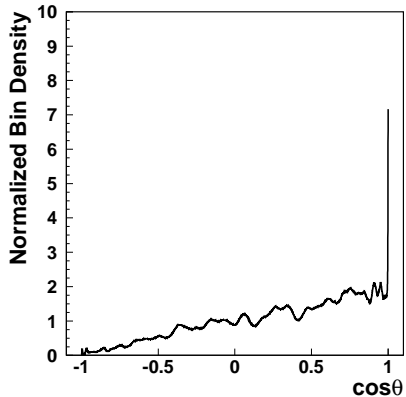}\\
\end{tabular}
\caption{The autocorrelation for the published AGASA events
\cite{Takeda:1999sg}.}
\label{fig:agasa_au}
\end{figure}

We also calculate the autocorrelation function for the published AGASA events
\cite{Takeda:1999sg}.
We show the result in figure~\ref{fig:agasa_au}. 
For this sample, we obtain 
$<\!\!\cos\theta\!\!>_{[0^\circ,10^\circ]}=0.99352$.

To study the relative sensitivity of AGASA and HiRes-I, we 
measure the value of $<\!\!\cos\theta\!\!>_{[0^\circ,10^\circ]}$ for multiple
simulated sets with a variable number of doublets inserted.  We then 
construct an interpolation of the mean value and standard deviation of
$<\!\!\cos\theta\!\!>_{[0^\circ,10^\circ]}$ from a given number of observed
doublets for each experiment.   This will allow us to state the number of
doublets required for each experiment in order for the 90\% confidence 
limit of 
$<\!\!\cos\theta\!\!>_{[0^\circ,10^\circ]}$ to be above the background value
of 0.99250.  In general, for a HiRes-I-like data set, the 90\% 
confidence lower limit corresponds to the mean expected background signal
with the inclusion of 6.25 doublets.  For an AGASA-like, the 90\% 
confidence lower limit corresponds to the mean expected background signal
with the inclusion of 5.5 doublets. 
This demonstrates that while AGASA has a slightly better ability to perceive 
autocorrelation, the sensitivity of the two experiments is comparable.
However, the observed HiRes-I signal corresponds
to the  90\% confidence upper limit with the inclusion of only 3.5 doublets
beyond random background coincidence.

\subsection{Conclusion}

We conclude that the HiRes-I monocular detector sees no 
evidence of clustering in its highest energy events.  Furthermore, the HiRes-I
monocular data has an intrinsic sensitivity to global autocorrelation such that
we can claim at the 90\% confidence level that there can be no more than 3.5 
doublets above that which would be expected by background coincidence
in the HiRes-I monocular data set above $10^{19.5}$~eV.  From this result,
we can then derive,
with a 90\% confidence level, that no more than 13\% of the observed HiRes-I
events could be sharing common arrival directions. This data set is
comparable to the sensitivity of the reported AGASA data set if one
assumes that there is indeed a 30\% energy scale difference between the two
experiments.  It should be 
emphasized that this conclusion pertains only to point sources of the sort
claimed by the AGASA collaboration.  Furthermore, because a
measure of autocorrelation makes no assumption of the underlying astrophysical
mechanism that results in clustering phenomena, we cannot claim that the HiRes 
monocular analysis and the AGASA analysis are inconsistent beyond a specified 
confidence level. 

\section{Acknowledgments}
This work is supported by US NSF grants PHY 9322298, PHY 9321949, 
PHY 9974537, PHY 0071069, PHY 0098826, PHY 0140688, PHY 0245428, PHY 0307098
by the DOE grant FG03-92ER40732,
and by the Australian Research Council. We gratefully
acknowledge the contributions from the technical staffs of our home
institutions. We gratefully acknowledge the contributions from the University
of Utah Center for High Performance Computing. The cooperation of 
Colonels E. Fisher and G. Harter, the US Army and the Dugway Proving Ground 
staff is appreciated.

\end{document}